# HISTORICAL NOTES

# Bhavnagar Telescope: the most widely travelled telescope in the country

N. Kameswara Rao, Christina Birdie and A. Vagiswari

*In the last decade of the 19th century Maharaja Takhtasingji Observatory was built at Poona (1888–1912) under the supervision of K. D. Naegamvala, with the grant from Maharaja of Bhavnagar (from where the name Bhavnagar Telescope must have originated). The story of this telescope from its inception to the current status is traced. Indian Institute of Astrophysics Archives has been extensively used to resource information for this note.*

Some telescopes are better known for the adventure and experiences they provided than the astronomy done with them. One such telescope is the Bhavnagar Telescope of Maharaja Takhtasingji Observatory, Poona (now non-existent). It spent more time sitting in boxes and travelling across the country than looking at the sky.

It started its career as a 16½ inch Newtonian telescope, one of the first few reflecting telescopes that was ever made[1]. It was a Grubb telescope manufactured by the famous Dublin telescope makers (see note 1). The order for the telescope was placed probably when K. D. Naegamvala from the College of Science, Poona, under University of Bombay visited Howard Grubb in July 1884 at Dublin. By 1885, Grubb apparently mentioned to Gill that he is already building a 16½ inch reflector for Bombay and also a Grubb spectrograph[1]. The choice of this odd size for the aperture of the telescope might have something to do with the availability of a 17-inch reflector tube which might be a part of the 'twin equatorial telescope' (a combination of reflector and refractor on the same mount) designed by Grubb. According to Glass 'The tube of this telescope may have been one of these which formed part of the "Twin Equatorial Telescope" at the Manchester Exhibition in 1887. It was then described as a 17-inch reflector, and since there is no record of such a telescope anywhere else, it can probably be identified with the Poona instrument.' Naegamvala[2] described the telescope as 'The principal instruments are a 16½ inch silver-on-glass Newtonian by Sir H. Grubb, with a 4-inch finder attached and a 6-inch equatorial refractor by elder Cooke... they are provided with several spectroscopes by Grubb, Hilger, Brown and others. …It is intended at present to restrict the work of the observatory to certain branches of spectroscopic research together with occasional observations of comets.' The telescope was not only used with it 'silver-on-glass' for observation but also 'with it silvering removed from the glass parabolic mirror' for observations of transit of Mercury during 9 May 1891 (ref. 3). We wonder where the next silvering was done to the mirror? Probably when it was sent back to Grubb.

'Before being shipped to India it along with the 250 pounds dome was inspected in 1887 or 1888 by Thomas Cushing, F.R.A.S. at India office's Lambeth observatory. The building at Poona was ready in 1888 but the telescope was not installed till the end of 1890' (ref. 4).

After 1894, the telescope took a different 'avatar', it became a Cassegrain telescope of 16½ inches. 'Some spectroscopic work of preliminary character was done during 1891, but it was found that the instrument used was altogether lacking in stability and was very weak in its driving parts. It was thereupon returned to Sir Howard Grubb for radical alterations' (Lockyer as quoted by Kochhar[4]). Probably the telescope was unable to take the weight of the spectrographs at Newtonian focus. When it came back from Grubb it became 'a Cassegrain reflector of 127-inches focus'. Usually it is not easy to convert a telescope from a Newtonian to a Cassegrain one because it requires the primary mirror to have a central hole in it and making a hole in the mirror after it has been figured is risky and could spoil the same. Fortunately the design of the Grubb photographic reflectors of that period had their primary mirrors with a central hole for viewing the focusing arrangement (on a screen at the top end)[1]. Existence of this hole in the design proved to be advantageous for later arrangements.

In the next 'avatar' the telescope had transformed into a 20-inch aperture one (Figure 1). Naegamvala paid another visit to Grubb during September 1896, probably to get the upgrade done by replacing the 16½ inch mirror with a 20-inch A. A. Common's mirror. Although there are no records of what transpired with Grubb, but in a letter dated 10 October 1896, addressed to W. H. M. Christie, the Astronomer Royal, Naegamvala states 'By April next the equipment of my observatory will consist of a 20″ Reflector 11′3″ focus with a six-inch guider mounted on a stand very nearly the same as that of Grubb's 13″ astrophotographic telescopes.' The telescope with its replaced mirror arrived in Poona some time in 1897. It certainly was available before January of 1901 as stated in a letter by Naegamvala to Christie[5,6].

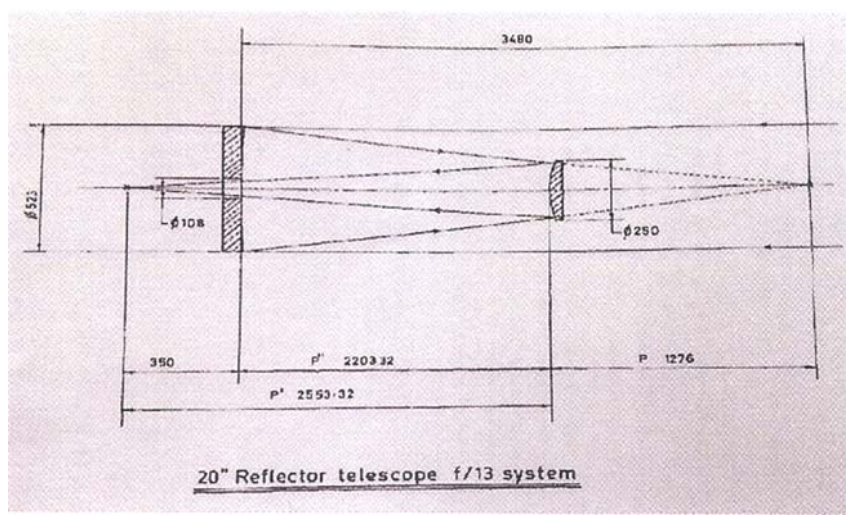

**Figure 1.** Optical layout of Bhavnagar Telescope (R. Rajamohan and J. P. A. Samson, private commun.).

All of the astronomy done (published) with the telescope during its life at Poona was accomplished when it was a 16½ inches telescope. A growth in size hardly helped in spite of several assertions by Naegamvala, who resolved to carryout observations, but it did not result in anything significant. Most of the observations reported during this period refer to photographing few nebulae from John Herschel's catalogue, e.g. M8, Nebula in Virgo. One of the first spectroscopic investigations with the telescope was related to the observation of [OIII] line of 5007 Å in Orion nebula, which confirmed that the line is sharp and narrow and not 'flauted' (faded as molecular bands) thus negating 'meteoritic hypothesis' as proposed earlier by Norman Lockyer[7]. It is strange that the spectroscopic observations of Nova 1901 Persius (= GK Per) obtained by Naegamvala were carefully observed with 'Vogel and Mclean spectroscopes of varying dispersions applied to the 16-inch guiding telescope of the photographic reflector of the observatory[5]' and not the 20-inch telescope (see note 2). Naegamvala observed P-Cygni type lines in the spectrum with emission and absorption for the H I Balmer lines and lines of He I. (Incidentally the focal length of the 16½ inch reflector was quoted differently at different places. Initially it was 127-inches[4] and in a letter written by Naegamvala dated 25 April 1895 (addressed to the Astronomer Royal) mentioned as 122-inches[8,9].)

After the retirement of Naegamvala, the Maharaja Takhtasingji Observatory was wound up and all its equipment, including the Bhavnagar Telescope were put in boxes and sent to Kodaikanal Observatory in 1912 by order of the Government of India. This widely travelled telescope is now in our possession for more than 100 years. The largest telescope in the country (at that time) securely remained in boxes for over three decades until A. K. Das[10] took interest and initiated its revival in 1949. The telescope was reassembled with missing parts being re-fabricated and erected by 1951 as part of the golden jubilee celebration of the Kodaikanal Observatory. The Kodaikanal Observatory was invited by the International Mars Committee to join the worldwide photographic and visual patrol of the 1954 opposition of Mars. The reactivated Bhavnagar Telescope was the main instrument that was used for these observations, which were published in Kodaikanal Observatory Bulletin No. 154 (ref. 11).

Soon after his arrival at Kodaikanal as Director, M. K. V. Bappu built a spectrograph for the telescope that formed the basis for several studies in stellar spectroscopy. Spectra of Wolf–Rayet stars, hot stars in associations, binary stars, Nova Delphini of 1967, comet Ikeya-Seki were some of the areas that were investigated using the telescope at Kodaikanal.

The telescope next arrived in Kavalur observing station in 1978 without its original mount and was substituted with a 'home-made mount'. It took another six years for the original mount to arrive at Kavalur. During these years spectroscopic work on few Be stars and Beta Cephei-type stars[12] was carried out.

During early 1980s proposals were made for a high-altitude national observatory to carry out infrared astronomy. The Department of Science and Technology, New Delhi funded a project to investigate sites in Ladhak, particularly around Leh for this purpose. J. C. Bhattacharyya, then Director of the Indian Institute of Astrophysics, Bangalore offered to loan the celebrated Bhavnagar Telescope to be used for site exploration as well as to perform certain astronomical studies. The telescope was installed at Sakar village, 3 km from Leh, by October 1984 (Figure 2). It remained there till November 1988, serving the needs of site evaluation from photoelectric observations of stars for estimating variations of atmospheric extinction, seeing, etc. for several years[13,14]. Few variable stars have also been monitored[15] during this time. Finally, it was realized that Leh is not the best place for optical studies (see ref. 16 for the final report of this experiment). The Bhavnagar Telescope has been brought back after its high-altitude mission and presently stays in boxes in Kodaikanal.

Bhavnagar Telescope remained the biggest telescope in the country from its arrival at Poona in 1888–90 till about 1968. It took different forms and configurations and remained the only spectroscopic telescope for night-time astronomy in the country for several decades. It travelled widely, at least twice to Dublin, after arrival in Poona, and then to Kodaikanal. From Kodaikanal it went Kavalur and then to Leh and back to Kodaikanal.

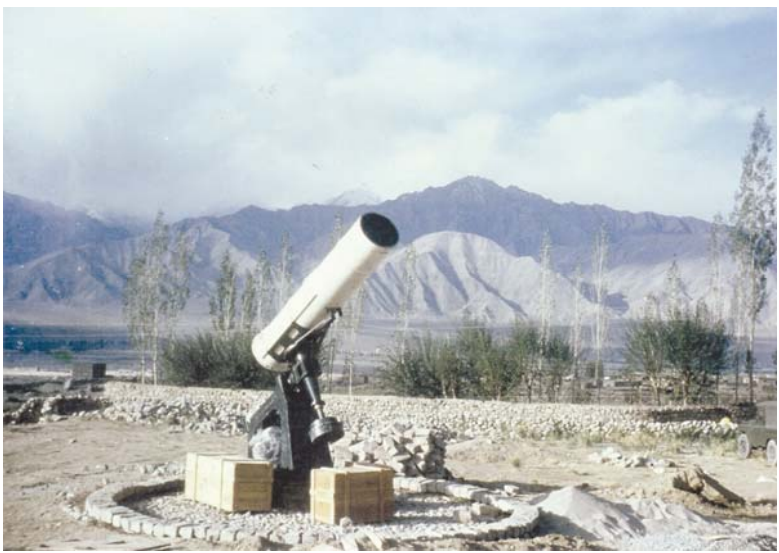

**Figure 2.** Bhavnagar Telescope at Leh.

The telescope constantly provided challenges to the observers. However, it did produce results, two Ph D theses and several papers have been published using this facility. It is hoped that the Bhavnagar Telescope, with such a colourful history, would be up and running again in the near future.

**Notes**

1. The, now defunct, 15-inch refractor and Carte de Ceil astrographic telescopes of Nizamiah Observatory were also made by Grubb.
2. See Anupama and Kantharia[17] for a recent study and an image of the remnant with 2-m Himalayan Chandra Telescope.

*N. Kameswara Rao, Christina Birdie\* and A. Vagiswari are in the Indian Institute of Astrophysics, Bangalore 560 034, India. \*e-mail: chris@iiap.res.in*